\documentstyle[twoside,fleqn,espcrc2]{article}

\newcommand{\AmS}{{\protect\the\textfont2
  A\kern-.1667em\lower.5ex\hbox{M}\kern-.125emS}}

\def\beq#1{\begin{equation}\label{#1}}
\def\beeq#1{\begin{eqnarray}\label{#1}}
\def\eeq{\end{equation}}
\def\eeeq{\end{eqnarray}}

\def\frac#1#2{ {{#1} \over {#2} }}

\def\rt){\right)}
\def\lt({\left(}
\def\rq]{\right]}
\def\lq[{\left[}

\title{Fracture Functions\thanks{Work done in collaboration with G. Veneziano.}
\thanks{Invited talk at the "QCD 94" Conference, Montpellier (France) July
7-13th 1994.}}
\author{L. Trentadue \address{INFN, Sezione di Roma II,
        and Dipartimento di Fisica, Universit\`a di Roma {\it "Tor Vergata"},\\
        via della Ricerca Scientifica, 1 , I-00133 Roma, Italy}
        \address{Dipartimento di Fisica, Universit\`a di Parma, 43100
          Parma, Italy}
        \address{present address: \\ Theory Division, CERN, Geneva,
Switzerland} }

\begin{document}

\begin{abstract}
We present a new approach to semi-inclusive hard processes in QCD by
means of $\it Fracture\;Functions$, hybrids between structure and fragmentation
functions. We briefly motivate and describe it together with
a list of possible applications.
\end{abstract}

\maketitle

\section{Introduction}

Asymptotic freedom \cite{af} together with general factorization theorems
\cite{fact} have
made possible to  predict a large variety of perturbative
hard processes \cite{alt}. In the QCD improved parton
model experimental cross-sections can be
computed by convoluting some uncalculable, but process
independent,  quantities with process-dependent, but calculable,
elementary cross-sections. For any given process, initiated by the hadrons $A$
and $B$:
$A+B\rightarrow A'+B'+...$, it is possible to write it in terms of
a pointlike, partonic cross-section $d\sigma$ convoluted with suitably defined
structure and fragmentation functions, $F_A^i$ and $D_j^{A'}$. The
mass-singularities
plaguing the radiatively corrected distributions, can be absorbed in the
structure and fragmentation functions. The cross-section
written in the factorized form is:

\begin{eqnarray}
\label{ga}
d\sigma(Q^2)= \sum_{k,...} \int dx_i\;dx_j..dz_k..F_A^i(x_i,Q^2)
\nonumber \\
F_B^j(x_j,Q^2)
\cdots d\sigma_{ij}^{k,...}(Q^2)\;D_k^{A'}(z_k,Q^2)...
\end{eqnarray}

where $d\sigma_{ij}^{k,...}(Q^2)$ is the pointlike partonic cross-section.
The universal structure and fragmentation functions $F_A^i(x,Q^2)$ and
$D_{i,A}(z,Q^2)$
do obey the evolution equations:

\begin{eqnarray}
\label{ev}
F_A^i(x,Q^2)=\sum_b \int_x^1 {dz \over z}
E_i^b(\frac {x}{z},Q^2,Q_0^2)\;\;\;\;\;\;\;\;\;\;\;\;\;
\\ \nonumber
\cdot F_A^b(z,Q_0^2)   \;\;\;\;\;\;\;\;\;\;\;\;\; \\ \nonumber
D_i^A(x,Q^2)=\sum_b \int_x^1 \frac{dz}{z} E_i^b ( \frac{x}{z}, Q^2, Q_0^2)
\;\;\;\;\;\;\;\;\;\;\;\;\;
\\ \nonumber
\cdot D_b^A(z,Q_0^2) \nonumber\;\;\;\;\;\;\;\;\;\;\;\;\;\;\;
\end{eqnarray}

Universality means that the structure functions, which can be measured in
deep inelastic lepton-hadron collisions, can be then  used to compute either
the same process or a completely
new hard reaction at a different scale.

\section{A new look to inclusive processes}

Let us reformulate, in a slightly different way, some well known
QCD results. Inclusive hard processes can be divided in two
classes: ${\it totally \; inclusive}$ and ${\it semi-inclusive}$
ones. Furthermore, within each of these classes, one can list
processes according to the number of the hadrons appearing in the initial
and in the final state $n_{i}$ and $n_{f}$ respectively.

We may consider a process with just
leptons in the initial state $n_{i}=0$
$e^+e^- \rightarrow hadrons$. If
no particular hadron is singled out in the final state then $n_{f}=0$. In
this case the process is completely calculable:
\beq{eetot}
 \sigma (e^+e^- \rightarrow H) = \Sigma _x  \sigma
(e^+e^-\rightarrow
 x)
\eeq
where, if $H$ is anything, the sum over $x$ runs over any partonic final state.
If $H$ represents three jets, $x$ will
be any number of partons in a three-jet configuration and so on.
The sum over the final
partons will eliminate all infrared and collinear singularities
and the cross-section will have a finite perturbative expansion
in $\alpha_s$ \cite{kln}\cite{kuv}.

Let us consider now  processes in which one hadron is present in
the initial state $n_i=1$, thus, typically, deep inelastic lepton hadron
scattering \cite{taylor}.
The  cross-section for  some hard process $H$:
 $\sigma(l+N \rightarrow l'+ H + X )$,
in which
no particular hadron is singled out in the hadronic final state $H+X$,
is well known to take the form \cite{alt}:

\beeq{distot}
\sigma_{l+N \rightarrow l'+H+X}=
\sum_j \int_0^1
\frac{dx}{x}
F_N^j(x,Q)\sigma_{H}^{j}(x,Q)
\eeeq

Where $Q$ is the virtuality of the photon
and $F_N^j(x,Q)$ is the structure function of the target $N$.
$H$ can represent any hard process as a jet, many jets, a photon and a jet, two
heavy quarks, etc...
The factorized form of eq.(\ref{distot})
 corresponds to  the fact that,
as result of the hardness of the collision, the final state consists
of two well separated clusters of particles, one (denoted by $X$)
originating from the target fragmentation and from the evolution
of the active parton and the other (denoted by $H$) coming from the
subsequent hard interaction of the active parton with the lepton.
In addition there will be some wee partons (or soft hadrons) which
cannot be unambiguously attributed  to either $H$ or $X$.
In the case of hadron-hadron hard collisions, the
cross-sections for $A+B \rightarrow H+X_A+X_B$
can be analogously factorized as:
\beeq{sigmas}
\sigma_{A+B \rightarrow H+X_A+X_B}=
\sum_{ij}\int \frac{dx_i}{x_i}\;\frac{dx_j}{x_j}\;\;\;\;\;\;\;\;\nonumber  \\
\cdot F_A^i(x_i, Q)\;\;
 F_B^j(x_j, Q)\;\;{\sigma}_{hard}^{i+j\rightarrow H}
(x_i,x_j;Q) \; .
\eeeq
Eq.(\ref{sigmas}) does not
contain, under this factorization hypotesis, new uncalculable quantities
besides
the ones we can
already measure
in deep inelastic scattering.

If a single hadron is detected in the final state $n_f=1$ then
the simplest case corresponds to the cross section:

\beq{ee}
\frac{d \sigma_{e^+e^- \rightarrow h+X}}{dz}=\sum_i
\sigma_{e^+e^-\rightarrow q_i \bar{q}_i}\;D_i^h(z,Q)
\eeq
which can be used to determine from the data the perturbatively
uncalculable fragmentation function $D_i^h(z,Q)$.
Thus, in this case, even processes with no initial hadron provide
important non-perturbative information.
The process $l+A\rightarrow l'+h+H+X$, according
to our previous discussion, will receive contributions
from two well separated kinematical
regions for the produced hadron $h$:
\beeq{curren}
\sigma_{l+A \rightarrow l'+h+H+X}= \sigma_{current}+
\sigma_{target} \nonumber \\
= \sigma_{l+A \rightarrow l'+(h+H')+X}
+\sigma_{l+A \rightarrow l'+H+(h+X')} \; .
\eeeq
For the first term, apart from the factor arising from target
structure function, no knowledge other than the one on
fragmentation
functions $D$ is needed. Such $"current"$ contribution has been widely
discussed in the literature \cite{ed} and we will not examine it here.
We shall instead concentrate on the second term claiming that its description
does require a new non-perturbative (but measurable) quantity, a
fragmentation-structure or "{\it fracture }" function \cite{tv}:
\beq{fra}
\sigma_{target}=\int_0^{1-z} \frac{dx}{x}
 M_{A,h}^i(z,x;Q)\;\sigma_{hard}^i(x,Q). \;
\eeq
This form clearly implies a new factorization
which will permit to describe the full target
fragmentation in terms of the single function $M$
without separating, as it is usually done, the
contributions of the active parton and that of the spectators \cite{feyn};
The factorized form in eq.(\ref{fra}) implies that, once $M$ is measured in
deep inelastic scattering no extra input is needed in order to compute
analogous quantitites in $hadron-hadron$ collisions.
Furthermore, it becomes possible to
introduce in QCD new uncalculable, but measurable and universal functions,
that we call "$fracture$ " functions telling us about the $structure$
function of a given target hadron
  once it has $fragmented$ (hence its name)
into another given final state hadron. Fracture
functions depend upon two
hadronic and one partonic label and on two momentum fractions,
a Bjorken $x$ and a Feynman $z$ variable:
\begin{eqnarray}
\label{Mdep}
 M=M_{p,h}^j(x,z;Q) \; .
\end{eqnarray}
One can also say that $M$ measures the parton distribution of the
object exchanged between the target  and the final hadron, without
making any model about what that object actually is. As for ordinary structure
functions,
the importance of measuring
such an object will be twofold: (1) it will teach us about the structure
of hadronic systems other than the usual targets, and (2) it can be
used as input for computing other hard semi-inclusive processes
at other machines, such as some future hadronic collider.
Furthermore, it has been recently observed in a next-to-leading evaluation of
single particle
cross-sections , that an entire class
of collinear divergences of hadrons emitted along initial state directions
are naturally absorbed by fracture functions \cite{dirk}.

\vskip .3 true cm

\noindent  {\bf 3. Properties of Fracture functions }
\vskip .1 true cm

In order to take into account the running of
$\alpha_s$, it is convenient to replace (see, e.g., Ref.\cite{kuv})
the evolution variable $Q^2$,  representing the hard scale of the
process, by the variable $Y$ defined by:
\beeq{y}
Y=\frac{1}{2\pi b}\ln\;[\frac{\alpha_s(\mu^2)}{\alpha_s(Q^2)}]
\eeeq
with $\mu$ the renormalization scale, $\alpha_s(Q^2)={(b\ln
\frac{Q^2}
{\Lambda^2})}^{-1}$ and the one loop $\beta$-function
coefficient  $b$   given by  $12\pi b=11N_C-2N_F$.
The evolution equation for the fracture function
$M_{p,h}^j(x,z;Q)$ feels the two distinct mechanisms
 of hadron production in the target
fragmentation region, the one coming
from the evolution of  the active   parton and the one
due to   fragmentation of the spectators. As a result, the evolution
equation for $M_{p,h}^j(x,z;Q)$ has two terms \cite{tv}:

\beeq{eveq}
\frac{\partial M^j_{p,h}(x,z;Y)}{\partial
Y}=\int_{\frac{x}{1-z}}^{1}
\frac{du}{u}\;
P_i^j(u)M_{p,h}^i(\frac{x}{u},z;Y)\;\;\;\;\;\;
\nonumber \\
+\int_{x}^{\frac{x}{x+z}}
\frac{u du }{x(1-u)}
{\hat{P}^{j,l}_{i}}(u)D_l^h(\frac{z u}{x(1-
u)},Y)\;\;\;\;\;\;\;\;\;\;\;\;\\ \nonumber
\cdot F_{p}^{i}(\frac{x}{u},Y)\;\;\;\;\;\;\;\;\;\;\;\;\;\;\;\;\;\;\;\;
\eeeq

 with $P_i^j(u)$ and $\hat{P}_i^{jl}(u)$ the regularized and real
\cite{kuv} Altarelli-Parisi vertices, respectively.

$D_l^h(z,Y)$ represents the
fragmentation function of the parton $l$ into the hadron $h$
and $F_p^i(x,Y)$
is the ordinary deep inelastic structure function.
$x$ and $z$ are the Bjorken variable of the
$i$-parton and the Feynman variable of the hadron-$h$.
It can be seen \cite{tv} that eq.(\ref{eveq}) has the solution:

\beeq{sol}
M^j_{p,h}(x,z;Y)=\int_{x}^{1-z}\frac{dw}{w}
E^j_i(\frac{x}{w},Y-y_0)\;\;\;\;\;\;\;\;\;\;\;\;\;\;\;\;
\nonumber \\
M_{p,h}^i(w,z;y_0)+\int_{y_0}^Ydy\int_{x+z}^1\frac{dw}{w^2}\int_{\frac{x}{w}}
^{1-\frac{z}
{w}}\frac{du}{u(1-u)}\;\;\;\;\;\;\;\;
\\
\cdot E_k^j(\frac{x}{wu},Y-y)
\hat{P}_i^{kl}(u)\;D_l^h(\frac{z}{w(1-u)},y)\;F_p^i(w,y).\;\;\;\;\;\;\;\;\;\;\;\; \nonumber
\eeeq

The first term takes the hadron distribution
at a given arbitrary scale $y_0$ and evolves it to the hard scale
$Y$ by means of the perturbative ``evolution'' function
$E^j_i(\frac{x}{x},Y-y_0)$ determined by
the evolution equation $\frac{\partial}{\partial y}E_i^j(x,y)=
\int_x^1\frac{du}{u}\;P_k^{j}(u)\;E_i^k(\frac{x}{u},y)$.
The second term describes the perturbative
evolution from $y_0$ to $Y$ of the  shower generated by
 the active parton. The shower
 generates perturbatively an inclusive distribution for the
 parton $l$ which finally fragments into
$h$. The second term in (\ref{sol})
contains $F$ and $D$ but not the fracture function $M$ itself.
 It can be also shown \cite{tv} that:
\begin{itemize}

\item
the solution given in eq. (\ref{sol})
does not depend on the arbitrary scale $y_0$, chosen as the starting point
of the evolution i.e.:

\beq{zero}
\frac{\partial}{\partial y_0}M^j_{p,h}(x,z;Y)=0 \; .
\eeq

\item
$M^j_{p,h}(x,z;Y)$ satisfies the natural momentum sum rule:

\beeq{summa}
\sum_h \int dz zM^j_{p,h}(x,z;Y)=(1-x)
\\ \nonumber
\cdot F_p^j(x,Y).\;\;\;\;\;\;\;\;
\eeeq

accounting for s-channel unitarity constraints.

\end{itemize}

\noindent  {\bf 4. Applications }
\vskip .1 true cm

Leaving to further work a more
detailed analysis, let us list  possible applications of fracture functions.
\begin{itemize}

\item
 One can simply consider $M_{N,h}^j(x,z;Q)$ for
large $z$ and  define that to be the structure function
of the leading trajectory that can be exchanged between the target
(here a nucleon) and the observed hadron. More generally, on the
basis of the Regge-Mueller analysis of inclusive cross-sections, we
may expect, as $z \rightarrow 1$, an expansion of the type:

\beeq{Regge}
M_{N,h}^j(x,z;Q)
\rightarrow \Sigma_{R} (1-z)^{1-2\alpha _R}
\\ \nonumber
\cdot F_R(\frac {x}{1-z};Q)\;\;\;\;\;\;\;\;
\eeeq

where the sum is over different Regge poles  of intercept
$\alpha_R$. $F_R(x;Q)$ may be defined to be the structure function of the
Rth Reggeon exchanged between $N$ and $h$. Such a parametrization can
be particularly suitable to describe diffractive processes recently observed
at the Tevatron and HERA \cite{diff}.

\item
One could compare fracture
functions for various quantum numbers of the $N-h$ system, and, in
particular, the relative amounts of valence
quarks, sea quarks and gluons in various channels.
Gluon-rich distributions for vacuum
quantum numbers ($N=h=p$), i.e. the so-called Pomeron structure
function \cite{pom}. At the opposite extreme, for
so-called exotic quantum numbers, we should find fracture functions which
are very rich in valence quarks. Examples of this type, with a proton
target, are $h=\bar p$ and $h= \bar K$ in which the fracture function
contains six and five quarks, respectively.

\item
 Various deep inelastic lepton-hadron processes can be used in order
to disentangle quark and gluon fracture functions for various $h$.
Thus, while the ordinary semi-inclusive cross-section can be used to
measure the quark distribution, production of heavy quarks can give
the gluon distribution. For $W$ and $Z$ production via the
Drell-Yan process, for example, it would be convenient to trigger on final
hadrons which give quark-rich
fracture functions. If, instead, one is interested in Higgs production
via the gluon-gluon fusion process, gluon-rich distributions will
have to be preferred.

\item
A possible application of fracture
functions is also to polarized lepton-hadron processes and to the question of
the so-called spin crisis \cite{spin}. The latter simply means that the
matrix element of the flavour-singlet axial current in the proton is
significantly smaller than one expects in the naive quark model.
Due to the $U(1)$ anomaly the spin problem
and the $U(1)$ problem are indeed related \cite{gt}.
 The question still remains of whether the smallness of the
proton spin is related really to the nature of the target
(the proton here)
or whether it is a more general (i.e. target independent)
property of the singlet axial current \cite{nsv}.
Fracture functions can provide new
informations by allowing one to measure matrix elements of the axial
currents in the $p+h$ state.

\end{itemize}

Fracture functions could open new
possibilities for studying hadron structure and for predicting
hard processes.
It would be interesting to see how they do compare with
real data of deep inelastic or hadron hadron
scattering.

\vskip .3 true cm
\noindent  {\bf Acknowledgements} \vskip .1 true cm

I would like to thank Gabriele Veneziano
for the stimulating and pleasant collaboration.
I would also like to thank Stephan Narison for
inviting me to this lively Conference.

This research is supported in part by the EEC Programme "Human Capital
and Mobility", Network "Physics at High Energy Colliders", contract
CHRX-CT93-0537 (DG 12 COMA).


\begin{thebibliography}{9}

%------------------------------------------------------------
%                 R E F E R E N C E S
%*****************************************************************
%--------------- 1
\bibitem{af}
        H. D. Politzer, Phys. Rep. 14 (1974) 129 and references therein.
 \bibitem{fact}
       D.~Amati, R.~Petronzio and G.~Veneziano, Nucl. Phys. B140 (1978) 54;
       R.K.~Ellis, H.~Georgi, M.~Mahaceck, H.D.~Politzer and G.G.~
       Ross, Phys. Lett. 78B (1978) 281; Nucl. Phys. B152 (1979) 285.
       J.~Collins, D.~Soper and G.~Sterman, in "Perturbative QCD",
       A.H.~Mueller Editor, World Scientific, 1989.

\bibitem{alt}
       G.\ Altarelli, Phys. Rep. 81 (1982) 1.

\bibitem{kln}
        T.~Kinoshita, Journ. of Math. Phys.,3,650 (1962); T.D. Lee and M.
Nauenberg,
Phys. Rev. 133 (1964) 1549;
        G. Sterman and S. Weinberg, Phys. Rev. Lett. 39 (1977) 1436;
        E. Fahri, Phys. Rev. Lett. 39 (1977) 1587; S.Catani, G. Turnock,
        B. R. Webber and L. Trentadue, Phys. Lett. 263B (1991) 491.

\bibitem{kuv}
       K.Konishi, A.Ukawa and G.Veneziano, Phys. Lett. 78B (1978) 243;
80B (1979) 259; Nucl. Phys. B157 (1979) 45.

\bibitem{taylor}
       R.\ Taylor, An Historical Review of Lepton Proton Scattering,
       SLAC-PUB-5832, June 1992, presented at the 1991 SLAC Summer
       Institute on Particle Physics: Lepton Hadron Scattering;

\bibitem{ed}
       E. Berger, Semi-Inclusive Inelastic Electron Scattering
        from Nuclei,
       Proceedings of the NPAS Workshop on Electronuclear
       Physics with Internal Targets, SLAC, 1987, SLAC-report 316,
       R. G. Arnold and R. C. Minehart Eds., p. 82.

\bibitem{tv}
       L.~Trentadue and G.~Veneziano, Phys. Lett. B323 (1994)201.

\bibitem{feyn}
         For a discussion of this and related arguments within the parton model
         framework see R.P.~Feynman, Photon-Hadron Interactions, W.A. Benjamin
Ed.,1972,
          lecture 55.

\bibitem{pom}
         P. V. Landshoff, The Structure of the Pomeron, presented at
         the 27th Rencontre de Moriond: Perturbative QCD and Hadronic
Interactions,
         March 22-28, 1992; Editions Frontieres, J. Tran Thanh  Van
         Ed., p. 393 and references therein.

\bibitem{diff}
 See  the talks of V. Del Duca and G. Levman at this Conference.

\bibitem{dirk}
          D. Graudenz, preprint CERN-TH.7300/94, June 1994.

\bibitem{spin}
 For a review, see, e.g., G. Veneziano, The spin
of the proton and the OZI limit of QCD, in "From Symmetries to
Strings: forty years of Rochester Conferences" (Okubofest, Ashok Das
Ed., WSPC 1990) p. 86; S. D. Bass and A. W. Thomas, The spin
structure of the nucleon, Adelaide preprints,
ADP-92-183/T115 (1992), ADP-93-218/T136 (1993).

\bibitem{nsv}
       S.~Narison, G.M.~Shore and G.~Veneziano, preprint CERN-TH.7223, April
1994.

\bibitem{gt}
  G. Veneziano, Mod. Phys. Lett. A4 (1989) 1605;
G.M. Shore and G. Veneziano, Phys. Lett. 244B (1990) 75, Nucl. Phys.
B381 (1992) 3; G.M. Shore, talk at this Conference.

\end{thebibliography}
\end{document}